\documentclass
[prl,twocolumn,superscriptaddress,floatfix,nobalancelastpage,showpacs]{revtex4-1}%
\usepackage{graphicx,bm,times}
\usepackage{amsmath}
\usepackage{amsfonts}
\usepackage{amssymb}
\usepackage{graphicx}

\begin{document}

\title{Field-induced nematic-like magnetic transition in an iron pnictide superconductor,
Ca$_{10}$(Pt$_{3}$As$_{8}$)((Fe$_{1-x}$Pt$_{x}$)$_{2}$As$_{2}$)$_{5}$}

\author{M. D. Watson}
\affiliation{Clarendon Laboratory, Department of Physics,
University of Oxford, Parks Road, Oxford OX1 3PU, U.K.}

\author{A. McCollam}
\affiliation{High Field Magnet Laboratory, Institute for Molecules and Materials, Radboud University, 6525 ED Nijmegen, The Netherlands}

\author{S. F. Blake}
\affiliation{Clarendon Laboratory, Department of Physics,
University of Oxford, Parks Road, Oxford OX1 3PU, U.K.}

\author{D. Vignolles}
\affiliation{Laboratoire National des Champs Magn\'{e}tiques Intenses (CNRS, INSA, UJF, UPS), Toulouse, France}

\author{L. Drigo}
\affiliation{Laboratoire National des Champs Magn\'{e}tiques Intenses (CNRS, INSA, UJF, UPS), Toulouse, France}

\author{I. I. Mazin}
\affiliation{Code 6393, Naval Research Laboratory, Washington,
D.C. 20375, USA}

\author{D. Guterding}
\affiliation{Institut f\"{u}r Theoretische Physik, Goethe-Universit\"{a}t Frankfurt, 60438 Frankfurt am Main, Germany}

\author{H. O. Jeschke}
\affiliation{Institut f\"{u}r Theoretische Physik, Goethe-Universit\"{a}t Frankfurt, 60438 Frankfurt am Main, Germany}

\author{R. Valent\'\i}
\affiliation{Institut f\"{u}r Theoretische Physik, Goethe-Universit\"{a}t Frankfurt, 60438 Frankfurt am Main, Germany}

\author{N. Ni}
\affiliation{Department of Physics and Astronomy
University of California, Los Angeles}
\affiliation{Los Alamos National Laboratory, Los Alamos, NM 87545}
\affiliation{Department of Chemistry, Princeton University, Princeton, NJ 08544}

\author{R. Cava}
\affiliation{Department of Chemistry, Princeton University, Princeton, NJ 08544}

\author{A. I. Coldea}\email[corresponding author:]{amalia.coldea@physics.ox.ac.uk}
\affiliation{Clarendon Laboratory, Department of Physics,
University of Oxford, Parks Road, Oxford OX1 3PU, U.K.}

\begin{abstract}

We report a high magnetic field study up to 55~T of the nearly
optimally doped iron-pnictide superconductor Ca$_{10}$(Pt$_{3}$As$_{8}$)((Fe$_{1-x}$Pt$_{x}$)$_{2}$As$_{2}$)$_{5}$ ($x$=0.078(6)) with a $T_{c} \approx$ 10 K using magnetic
torque, tunnel diode oscillator technique and transport measurements. We
determine the superconducting phase diagram, revealing an anisotropy of the
irreversibility field up to a factor of 10 near $T_{c}$ and signatures
of multiband superconductivity. Unexpectedly, we find a spin-flop like anomaly
in magnetic torque at 22~T, when the magnetic field is applied perpendicular to the
 ({\it ab}) planes, which becomes significantly more pronounced as the temperature is lowered to 0.33~K.
As our superconducting sample lies well outside
the antiferromagnetic region of the phase diagram, the observed field-induced transition
in torque indicates a spin-flop transition \textit{not of long-range ordered moments},
but of nematic-like antiferromagnetic fluctuations.

\end{abstract}
\date{\today}
\maketitle


It is widely believed that superconductivity in iron pnictides and selenides is
intimately related to their magnetic properties. The unusual stripe order \cite{Lumsden2010}
which breaks the four-fold symmetry (or, more precisely, fluctuations associated with
this order) are often considered to play a key role in superconducting
pairing. According to this point of view, short-range antiferromagnetic fluctuations
survive well above the N\'{e}el temperature and gain additional stability by
acquiring a {\it nematic} component, that is,
 the simultaneous excitation of antiferromagnetic fluctuations with the same wave
vector but different phases leads to orbital fluctuations and
additional energy gain. Detecting such {nematic} fluctuations above the structural transition is
extremely challenging, although there is some experimental evidence to their
existence \cite{Kasahara2012a, Shimojima, Chu2012a,Fernandes2010}. In this letter we present
a magnetic-field tuned experiment that suggests the existence of nematic-like antiferromagnetic fluctuations
with preferential spin directions in an nearly optimally doped Fe-based superconductor
in the absence of a true long-range magnetic order.

\begin{figure}[th]
\centering
\includegraphics[width=7cm]{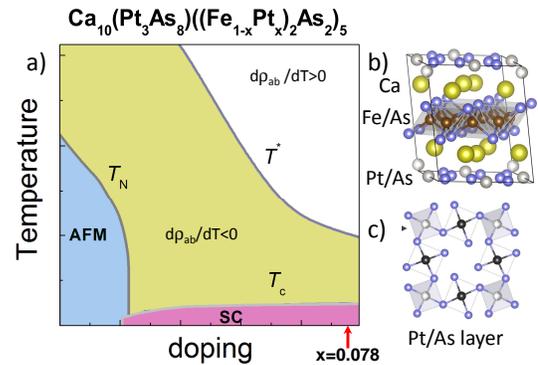}
\caption{a) Schematic phase diagram of Ca$_{10}$(Pt$_{3}$As$_{8}$)((Fe$_{1-x}$Pt$_{x}
$)$_{2}$As$_{2}$)$_{5}$ as a function of electron doping $x$
 based on Refs. \cite{Ni2011a,Xiang}. Our compound lies at the right of
 this phase diagram as indicated by the arrow.  (b) The structure of the 10-3-8 phase
 in which the Fe-As layers, Ca atoms and  Pt-As layers (in (c)) form a
layered structure that crystallizes in the low-symmetry $P\bar{1}$
triclinic space group.}
\label{fig:1038struct}
\end{figure}

Recently, Ca$_{10}$(Pt$_{3}$As$_{8}$)(Fe$_{2}$As$_{2}$)$_{5}$ was found to be
the parent compound of a new class of Fe-based superconductors, commonly
called 10-3-8 \cite{Ni2011a}. As shown in Fig.~\ref{fig:1038struct}(b) and (c), the
characteristic Fe-As layers  are separated by Ca atoms and a Pt$_{3}$As$_{8}$ plane and the phase diagram for this system
  shows similar features to those found in other families of iron-based superconductors.
Recently, optical imaging\cite{Cho2012}, NMR\cite{Zhou2013}, powder x-ray diffraction and $\mu$SR measurements \cite{Sturzer2013} have indicated the structural / magnetic phase transitions occur in the parent 10-3-8 compound around 100 K, where the NMR measurement probing the $^{75}$As environment suggests that the Fe-As planes have a striped antiferromagnetic order similar to BaFe$_2$As$_2$. Superconductivity occurs under applied pressure \cite{Gao2013} or with
electron doping, either by La doping onto the Ca site \cite{Ni2013}, or Pt
doping onto the Fe site (Fig.~\ref{fig:1038struct} (a)).
As the symmetries of the Fe-As and Pt-As planes are incompatible in the 10-3-8
phase, the system crystallizes in the triclinic $P\bar{1}$ group\cite{Ni2011a}.
 However, given the large separation of
the two subsystems the interaction of Fe-As with the Pt-As layer is weak and
thus the electronic properties of the Fe-As layer are expected to follow the
tetragonal symmetry as found in other iron-based superconductors.  This has
been confirmed by angle-resolved photo-emission spectroscopy
\cite{Neupane2012,Thirupathaiah2013}.

\begin{figure}[t]
\centering
\includegraphics[width=8.5cm]{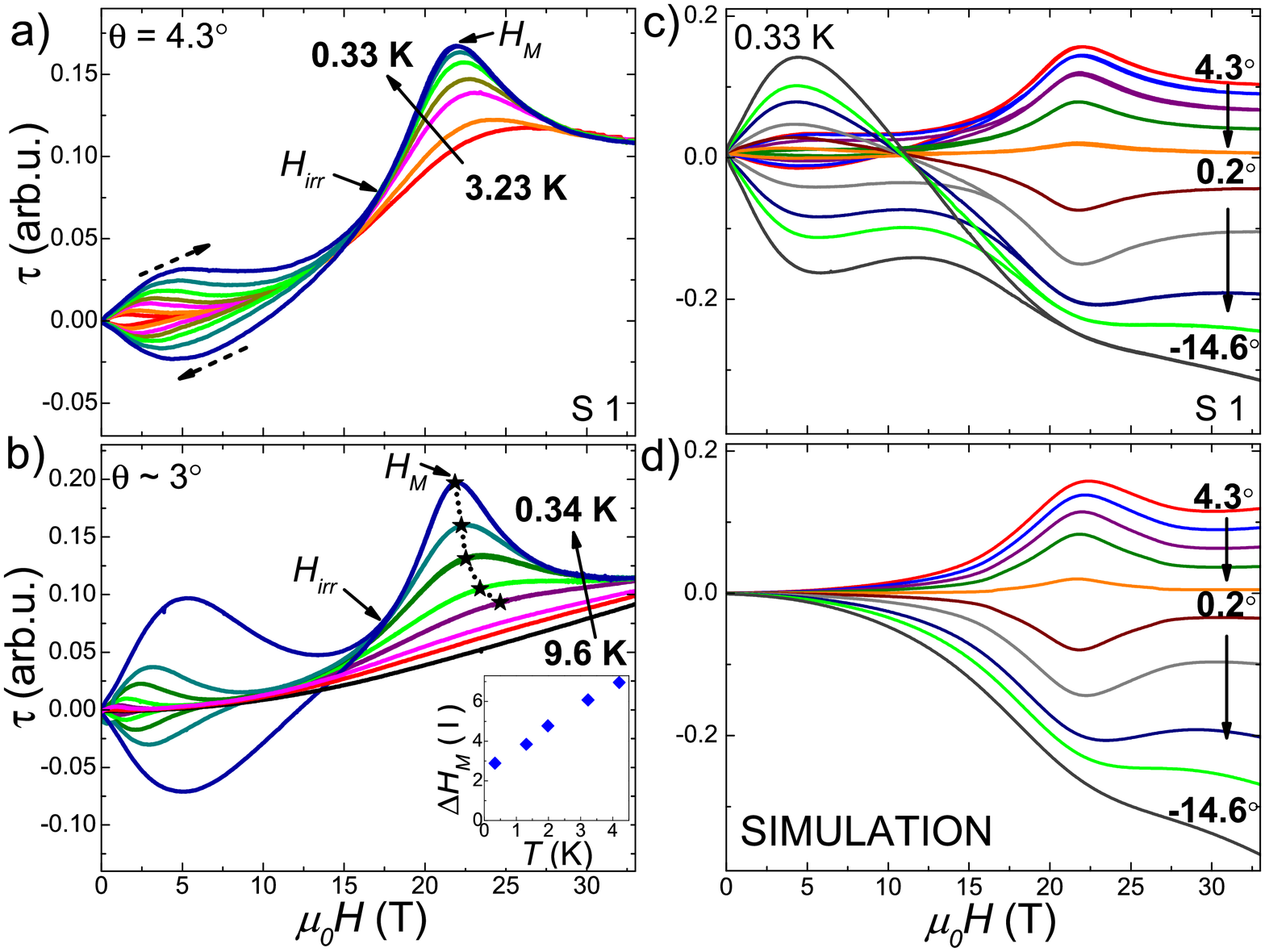}
\caption{a) Magnetic torque in an applied magnetic field at constant temperatures
between 0.33~K and 3.23~K for
a 10-3-8 sample S1 when $H{} \perp{} ab$ (within $\theta=4.3^{\circ}$, where $\theta$ is the angle
between the magnetic field and the normal to the {\it ab} planes). Solid arrows indicate
the positions of the irreversibility field $\mu_{0}H_{irr}$ and the peak of
the magnetic anomaly at $\mu_{0}H_{M}$, dashed arrows show field sweep direction. b) Magnetic torque versus field on
sample S2 up to 9.6~K. Stars indicate the position of the maxima
at $\mu_{0}H_{M}$ determined after subtraction of the 9.6~K signal. Inset: the
temperature dependence of  the peak width, $\Delta H_{M}$, fitted to a
Lorentzian. c) Field dependence of torque as function of angle
$\theta$ at 0.33~K. d) A simulation of torque based
on a spin-flop and a paramagnetic contribution described below,
assuming the magnetic moments are aligned perpendicular to the {\it ab}
plane in zero field.}
\label{fig:TorqueFigure}
\end{figure}

In this letter we report high magnetic field studies up to 55~T of the nearly
optimally-doped superconducting Ca$_{10}$(Pt$_{3}$As$_{8}$)((Fe$_{0.922}%
$Pt$_{0.078}$)$_{2}$As$_{2}$)$_{5}$, with $T_{c}\approx$ 10 K, using magnetic
torque, tunnel diode oscillator technique and transport measurements. We
determine the superconducting phase diagram, finding a high
anisotropy up to $\gamma_{H} = 10$ near $T_{c}$ and an irreversibility field
up to $32$ T. Furthermore, we find an anomaly at low
temperatures at 22~T that is consistent with a spin-flop transition,
when the magnetic field is perpendicular to the Fe-As planes
and in close proximity to the
superconducting phase. We suggest that
this transition - which occurs
in the absence of any long-range magnetic order in a quasi-two dimensional system -
could originate from strong nematic-like antiferromagnetic fluctuations
between two configurations with different preferential
orientations of the Fe spins.


Sample growth details are presented elsewhere \cite{Ni2011a}. X-Ray measurements confirmed the $P\bar{1}$ triclinic symmetry group
\footnote{The lattice parameters are $a$=8.7851(13){\AA }, $b$%
=8.7852(12){\AA }, $c$=10.6745(17){\AA }, $\alpha$ =94.66(1), $\beta
$=104.23(1), $\gamma$=89.91(1), slightly different from those found for the
parent compound reported in Ref. \onlinecite{Ni2011a}}.
Wavelength-dispersive X-ray measurements give a doping level of
$x = 0.078(6)$ by constraining Fe+Pt=13 in the chemical formula.
\footnote{WDX was performed on several areas of a mm-sized crystal from the same batch as
the samples in which the anomaly was observed. Our value of $x=0.078(6)$ is found by constraining
N(Fe+Pt)=13. The point-to-point variation was on the order of the measurement uncertainty.}
Transport measurements were performed in a Quantum Design PPMS in magnetic
fields up to 14~T and also in pulsed fields up to 55~T at LNCMI, Toulouse with
low-resistance electrical contacts made using Sn solder. Torque measurements
were performed at low temperatures (down to 0.33~K) in magnetic fields up to
33~T at the HMFL in Nijmegen and in pulsed fields up to 55~T at LNCMI in
Toulouse. Single crystals with typical size $\approx$
150$\times$150$\times$30 $\mu{}$m were used for torque measurements using
highly sensitive piezo-resistive microcantilevers. All samples measured from
the batch were found to be superconducting.


Fig.~\ref{fig:TorqueFigure} shows magnetic torque data on two crystals (S1 and S2) when the
magnetic field is approximately perpendicular to the {\it ab} planes.
The typical hysteresis that occurs in the superconducting state
due to the pinning of the flux vortices allows the determination of the
superconducting phase diagram from the position of the irreversibility field $\mu_{0}H_{irr}$, as
indicated in  Fig.~\ref{fig:TorqueFigure}(a) and (b). Furthermore, at
magnetic fields above the superconducting hysteresis, we detect a clear
anomaly at 22~T in the magnetic torque data that becomes sharper as the temperature is lowered to 0.33 K.
This effect was reproduced in all the crystals
investigated and no further anomalies were detected up to 55~T. This anomaly
is well-defined only when the applied field is nearly perpendicular to the {\it ab} plane, and it is symmetric around this
direction, as shown in Fig.\ref{fig:TorqueFigure}(c). The width of this
peak at $\mu_{0} H_{M}$ (the maximum in the torque signal after
subtracting the featureless 9.6 K sweep), quantified by the
half-width of Lorentzian fit to the subtracted data (inset of
Fig.~\ref{fig:TorqueFigure}(b)), shows a linear decrease with temperature suggestive
of slowing down of magnetic fluctuations.
The observed behavior of magnetic torque in 10-3-8 is rather unusual and
is in contrast to the paramagnetic $\tau{}\sim H^{2}$ background torque typically
observed in other iron-based superconductors above the irreversibility field
\cite{Coldea2008,Putzke2012}. The observed anomaly at 22~T is likely to be of
magnetic origin and the large magnitude of the anomaly compared to the
superconducting hysteresis suggests that a significant proportion of the
magnetic ions are involved, so an explanation involving impurities is
unlikely.

In order to understand this behavior we consider first the Pt ions. Heavy Pt ions have strong spin-orbit coupling, and may experience large magnetic anisotropy.
 In the doped material, there are two distinct types of Pt: one in the Fe plane (Pt$^{4+})$ and another one in the Pt-As layer (Pt$^{2+})$
 \footnote{The square-planar coordination of Pt in the Pt-As layer creates a strong crystal field, with a high antibonding $d_{x^{2}-y^{2}}$ band, and all other $d$-states occupied, so the valency of the interlayer Pt is ${2}^{+}$, as opposed to the Pt$^{4+}$ substituting for Fe. The arsenics in the Pt$_{3}$As$_{8}$ plane form covalently bound dimers with an effective valency of ${4}^{-}$ per dimer so that the ionic electron count is
 Ca$_{10}^{2+}$ Pt$_{3}^{2+}$(As$_{2}$)$_{4}^{4-}${Fe}$_{10}^{2+}${As}$_{10}^{3-}$.}.
We do not expect the latter to be magnetic, given the {\it d}$^{8}$ configuration of Pt$^{2+}$, but the former, on the first glance, might spin polarize. Even with a relatively weak polarization it would have provided for a measurable anisotropy of the magnetic response. For that reason, we performed density functional theory (DFT) calculations on a $\sqrt{5}\times \sqrt{5}$ 10-3-8 supercell where one Fe was substituted by Pt (see Supplementary Material, SM) and looked for spin polarization on Pt. We performed full lattice relaxation to avoid applying artificial internal pressure to the in-plane Pt. However, for both Pt positions (in-plane and intermediate layer) our calculations render completely polarization-free Pt, even though the surrounding Fe atoms acquire full polarization. Adding Hubbard $U$=2.5~eV and $J$=0.5~eV does not create spin polarization on Pt either.

Another way to rationalize this result is to look for spin-flop-like physics in the Fe subsystem. Although there is no reported long-range
antiferromagnetic order at this doping (see Fig.\ref{fig:1038struct}),
there are certainly antiferromagnetic fluctuations, which could become sufficiently long-lived at low temperatures to contribute to the magnetic torque. At a temperature $T\lesssim K,$ where $K$ is the single-ion magnetic anisotropy energy, the fluctuating spins at zero field will lie predominantly along the easy axis and may be driven into a spin-flop transition by an external field; this will  manifest in magnetic torque as a peak at $H_{SF}$, when the magnetic field is aligned along the zero-field easy axis. In experiments (Fig. \ref{fig:TorqueFigure}(a,b)) the sharpness of the anomaly at $H_M$, which we now associate with a spin-flop field $H_{SF}$, varies substantially between 0.33~K and 3.23~K, and is completely featureless by $\sim$10 K. Therefore a qualitative association of the temperature and magnetic anisotropy energy scales would suggest that $K$ is of the order of $\sim$1 K (86 $\mu$eV). A direct calculation of $K$ from DFT for the Ca-10-3-8 \textit{parent} compound ($x$=0) gives a value of $K\approx 106$ $\mu$eV  which is of the same order of magnitude as the measured magnetic energy scale
 but predicts $b$ as the easy axis. In our experimental geometry the spin-flop explanation
 requires the $c^{\ast}$ axis, perpendicular to the {\it ab} planes, to be the easy axis. 
 However, analysis
of the magnetic anisotropy in other pnictide compounds shows that DFT easy axis predictions does not
always agree with experimental observations (see SM).
A spin-flop field $H_{SF}$ could be linked to {\it ab-initio} calculations by
 the exchange energy that is calculated using the difference, $\Delta$, between the ferro- and antiferromagnetic
ordered states, derived as $\mu_{0}H_{SF}=2\sqrt{K\Delta}/M,$ where $M$ is the Fe moment
(see SM). 
However, in Ca-10-3-8 the ferromagnetic state converges only in a fixed
moment calculation which results in a large overestimation of $\Delta$ and correspondingly, $H_{SF}$.
Comparison of $H_{SF}$ calculated for Ca-10-3-8 and other iron-based superconductors shows that
DFT overestimates this value (see SM).

We can simulate the angular dependence of torque in a spin-flop scenario,
including a Lorentzian broadening term ($\Gamma \approx 2.1$~T) and adding an
additional $\tau\sim H^{2}\times{}\sin{2\theta}$ term to account for a
 paramagnetic contribution, as described in the SM. While a magnetization measurement of a spin-flop transition usually
shows a magnetization step or slope change at the
spin-flop field, magnetic torque as a function of applied field gives rise to
a peak  if the field is applied {\it parallel to the easy
axis} \cite{Kawamoto2008}. Despite the simplicity of the model, the correspondence with the
experimental data  above $\mu_0 H_{irr}$ is remarkable, as shown in Fig.~\ref{fig:TorqueFigure}(d).
The model assumes that the (fluctuating) magnetic moments are aligned
perpendicular to the {\it ab} plane in contrast to the in-plane
collinear antiferromagnetic order typically found in the parent compounds of
the iron-based superconductors. However, the anisotropy energies are not large
(see SM) and given the more unusual crystal structure of 10-3-8, it is possible that the easy axis is perpendicular to the {\it ab} planes.
 One may ask why similar spin-flop transitions have not been observed in other
Fe pnictides. On this point, it is important to keep in mind that the magnetic anisotropy
in these materials is not universal: neutron scattering studies suggest that in LaFeAsO,
SrFe$_{2}$As$_{2}$ and BaFe$_{2}$As$_{2}$ the spins are aligned in-plane \cite{Lumsden2010}, but in NdFeAsO \cite{Marcinkova} along $c$. In the latter case the spin flop may be masked by spin-reorientation transitions
in the Nd subsystem.
Additionally, there is evidence in polarized neutron scattering measurements
that the anisotropy of low-energy spin fluctuations~\cite{Liu2012b} in electron-doped Ba(Fe$_{1-x}$Ni$_{x}$)$_2$As$_2$
does not reflect the easy axis of the ordered phase.

\begin{figure}[ptb]
\centering
\includegraphics[width=8.5cm]{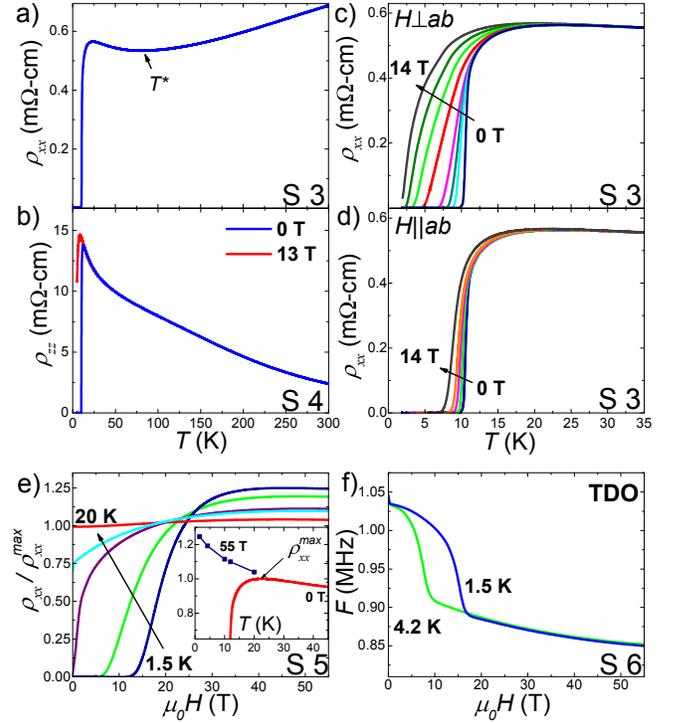}
\caption{Transport
measurements on single crystals of 10-3-8. (a)
Temperature dependence of in-plane resistivity in zero field. The arrow
indicates the position of the minimum in resistance at $T^\ast$. (b) The
inter-plane resistance that increases significantly with reducing
temperature. (c) Temperature dependence of in-plane resistivity is strongly
suppressed and broadened when $H \perp ab$ planes. (d) The in-plane resistivity show little
variation when $H \parallel ab$ planes. (e) Resistivity
measurements in a pulsed magnetic field. The data is renormalized to the value of the sample
at $\rho_{max}$ (indicated by the arrow in inset).
Inset: The normal resistance at 55~T as function of temperature plotted together with the zero-field
resistivity. (f) TDO measurements as function of magnetic field
showing a superconducting transition but no further anomaly at 22~T.}%
\label{fig:1038_Transport_Summary}%
\end{figure}

We have also performed in-plane and out-of-plane transport measurements to
characterize the superconducting phase and to determine whether the
transition in applied field can be detected in transport. Fig.~\ref{fig:1038_Transport_Summary}(a)
shows the temperature dependence of
the in-plane resistance that has a crossover from a metallic-like
to semiconductor-like regime \cite{Xiang} at a temperature $T^\ast$ around
80~K before it becomes superconducting at $T_{c}\approx$ 9.7 K. The application of a magnetic field perpendicular to the planes (Fig.~\ref{fig:1038_Transport_Summary}(c)) causes a substantial broadening of the superconducting transition. However, the
superconducting properties are highly anisotropic, and the application of a
field parallel to the {\it ab} planes in Fig.~\ref{fig:1038_Transport_Summary}(d)
does not suppress the superconducting transition to the same extent.
This broadening indicates strong thermal fluctuations of the vortex lattice, which
can be quantified by the Ginzburg number $Gi
\approx0.16$ \cite{Kim2012}, significantly higher than typical pnictides
(using a large penetration depth of $\lambda_{ab}(0)\approx1000$ nm\cite{Kim2012} and combined with the measured upper
critical field). In contrast with the in-plane measurements, the inter-plane resistance (Fig.~\ref{fig:1038_Transport_Summary}(b)) is found to increase with decreasing temperature by a factor $\sim$~6 suggesting strongly incoherent transport between the planes.
In-plane resistivity measurements (Fig.~\ref{fig:1038_Transport_Summary}(e)) and penetration depth measurements \cite{Drigo2010a} (TDO in Fig.~\ref{fig:1038_Transport_Summary}(f)) in very high magnetic fields up to 55~T
show no anomalies that could be associated with the magnetic transition (when $H\perp{}ab$),
 beside the expected transition from superconducting to the normal state.
By suppressing the superconducting state by the
application of a magnetic field of 55~T, the normal state resistance tends to
increase exponentially at low temperatures suggesting
field-induced charge localization in 10-3-8, as shown in the inset of Fig.~\ref{fig:1038_Transport_Summary}(e).

\begin{figure}[ptb]
\centering
\includegraphics[width=6cm]{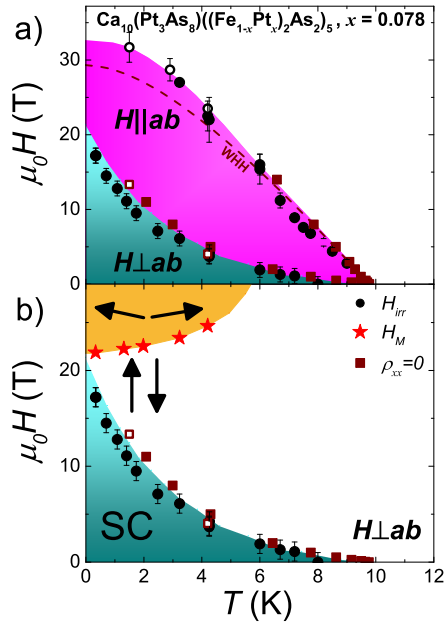}
\caption{(a) The superconducting field-temperature phase diagram of nearly optimally
doped Ca$_{10}$(Pt$_{3}$As$_{8}$)((Fe/Pt)$_{2}$As$_{2}$)$_{5}$ for two
different orientations with respect to the ({\it ab} plane), as obtained
from magnetic torque (circles) and transport measurements (squares). Open
symbols indicate the pulsed fields measurements. Dashed line is the WHH
fit to $H||ab$ data; the coloured phase boundaries are a guide to the eye. (b) The
magnetic field-temperature phase diagram for $H\perp{}ab$ showing the position
of the magnetic anomaly at $H_{M}$ (stars - taken from Fig.~\ref{fig:TorqueFigure}b)
 and the irreversibility field, $H_{irr}$ in the superconducting (SC) state.
 Black arrows represent the energetically
favorable orientations of the fluctuating antiferromagnetic spins.}%
\label{fig:PhaseDiagram}%
\end{figure}

Based on the transport and torque data we have constructed  the phase
diagram of the superconducting
Ca$_{10}$(Pt$_{3}$As$_{8}$)((Fe$_{1-x}$Pt$_{x}$)$_{2}$As$_{2}$)$_{5}$ ($x=0.078$),
as shown in Fig.~\ref{fig:PhaseDiagram}(a).
Near $T_{c}$, the anisotropy parameter $\gamma_{H}={H_{c}^{||ab}(T)}%
/{H_{c}^{\perp{}ab}(T)}$ has a value of 10, similar to the 1111 iron pnictides
\cite{Carrington2009} but significantly larger than the value of $\sim$2 in the 122s \cite{Yuan2009a}.
When the magnetic field is parallel to the
conducting plane, we use the single-band Werthamer-Helfand-Hohenberg (WHH) model \cite{WHH1966} and fit the zero-resistance
data points near $T_{c}$ to predict $H_{c2}(0)\approx{}$ 30 T, which is close to
the irreversibility field of 32~T measured in torque at 1.5~K. When the
magnetic field is perpendicular to the Fe-As layers ($H\perp{}ab$), the
superconducting phase boundary shows a strong concave curvature that cannot be
captured by the one-gap WHH model and is suggestive of multi-band superconductivity
in this system, as seen in the closely related 10-4-8 superconducting phase \cite{Mun2012} or
other anisotropic systems such as the 1111 pnictides \cite{Jaroszynski2008}
and cuprates.

While the superconducting phase diagram of the 10-3-8
has features found in other anisotropic superconductors,
the magnetic region induced by the high magnetic fields
at low temperatures is rather unusual.
The phase diagram under pressure \cite{Gao2013}, which is similar overall to that as function of doping shown in Fig.~\ref{fig:1038struct} (a),
tracks the minimum in the resistivity at
a temperature $T^\ast$  well above the structural transition,
suggesting that this temperature
could be linked to the gradual appearance of nematic fluctuations
(and the corresponding loss of the carrier density).
Due to the more strongly two-dimensional
character of the 10-3-8 compared
 to other Fe-pnictides (for instance, the long interlayer stacking distance of 10.27\AA{}
 and the lower N\'{e}el temperature in the parent
compound suggest weaker interlayer coupling), nematic antiferromagnetic spin fluctuations in the Fe-As planes without 3D long-range order are likely to exist over a large area of the phase diagram, roughly characterized by the resistivity upturn at $T^\ast$ (Fig. \ref{fig:1038struct}). It is these antiferromagnetic fluctuations along the $c$-axis that
could be responsible for the observed spin-flop transition in torque in an applied magnetic field.

In conclusion, we have used high magnetic fields to map out the superconducting phase diagram of Ca$_{10}$Pt$_{3}$As$_{8}$)((Fe$_{0.922}$Pt$_{0.078}$)$_{2}$As$_{2}$)$_{5}$ finding a high anisotropy up to $\gamma_{H}=10$, and an irreversibility field up to $\approx $ 32 T for $H||ab$. Most importantly, we reveal a field-induced magnetic transition in torque measurements at 22~T at low temperature not previously observed in Fe-based superconductors, which we attribute to a spin-flop of antiferromagnetic fluctuations in the Fe-As plane.
This thermodynamic measurement at low temperature demonstrates that the magnetic field is an effective tuning parameter
of these antiferromagnetic fluctuations which may be responsible for the pairing in the 10-3-8 iron-based superconductors. These fluctuations are likely to be nematic fluctuations that reflect the breaking of the local tetragonal symmetry in the Fe-As plane in 10-3-8 over a large range of temperature and doping. Further work to quantify the nature of these magnetic fluctuations needs to be provided by other experimental techniques, such  as polarized neutron scattering experiments.


\begin{acknowledgments}
We acknowledge fruitful discussions with Andrew Boothroyd and Susie Speller for technical support. This work was supported by EPSRC (EP/I004475/1) and part of the work by the EuroMagNET II (EU Contract No. 228043). AIC acknowledges an EPSRC Career Acceleration Fellowship. DG, HOJ and RV acknowledge support from the DFG through grant SPP1458. NN acknowledge support from UCLA, Marie Curie fellowship (LANL) and AFOSR MURI on superconductivity. D.G. acknowledges support
from the German National Academic Foundation. I.I.M. acknowledges
support from the Funding from the Office
of Naval Research (ONR) through the Naval Research
Laboratoryʼs Basic Research Program, and from the Alexander
von Humboldt Foundation.

\end{acknowledgments}






%

\end{document}